\documentclass [prl,amsmath,showpacs,twocolumn,preprintnumbers,superscriptaddress]{revtex4-1}

\usepackage{mathptmx}
\usepackage{microtype}

\usepackage{graphicx}
\usepackage{dcolumn}
\usepackage{amsmath}
\usepackage{mathtools}
\usepackage{bm}
\usepackage{amssymb}

\renewcommand*{\d}[1]{\mathrm{d}#1}

\newcommand*{\dalembert}{\square}

\newcommand*{\grad}{\hm{\nabla}}
\newcommand*{\bdot}{\bm\cdot}

\begin{document}

\title{Photon Orbital Angular Momentum and Proca effects in rotating and charged spacetimes}

\author{F. Tamburini} 
\address{Department of Astronomy, University of Padova, vicolo dell' Osservatorio 3, Padova, Italy}

\author{B.\,Thid\'e}
\address{ Swedish Institute of Space Physics,  Physics in Space,  {\AA}ngstr\"{o}m Laboratory,
 P.\,O.~Box~537,  SE-751\,21, Sweden}
 
\begin{abstract}
We analyze the effect of Proca mass and orbital angular momentum of photons imposed by a structured plasma in Kerr-Newman and Reissner-Nordstr\"om-de Sitter spacetimes. The presence of characteristic lengths in a turbulent plasma converts the virtual Proca photon mass on orbital angular momentum, with the result of decreasing the virtual photon mass. The combination of this plasma effect and that of the gravitational field leads to a new astrophysical phenomenon that imprints a specific distribution of orbital angular momentum into different frequencies of the light emitted from the neighborhood of such a black hole. The determination of the orbital angular momentum spectrum of the radiation in different frequency bands leads to a complete characterization of the electrostatic and gravitational field of the black hole and of the plasma turbulence, with fundamental astrophysical and cosmological implications.

\end {abstract}


\maketitle

\section{Introduction}

One of the most exotic predictions of EinsteinÕs Theory of General Relativity are black holes (BH) endowed with electric charge and consequently with an electrostatic field \cite{1992mtbh.book.....C}. In the general astrophysical scenario, black holes are expected to be surrounded by a turbulent plasma and,
through a selective capture of charged particles \cite{Dolgov200797}, to generate their electrostatic field that is described in the simplest scenario by the Reissner-Nordstr\"om  metric \cite{PhysRevD.66.024010}. 

When the source of the gravitational field is rotating, the spacetime behavior changes dramatically, and the Reissner-Nordstr\"om metric is replaced by the Kerr-Newman metric.  In addition to the effects of the electrostatic field, because of the rotation, the norm of the timelike Killing field is positive in the ergosphere that extends outside the BH horizon. The asymptotic time translation Killing field, $\xi^\alpha=(\partial/\partial t)^\alpha$, becomes spacelike and an observer in the ergosphere cannot remain stationary even if it is orbiting outside of the BH; observers inside the ergosphere are forced to move in the rotation direction of the BH following the lines of field of the electrostatic field. 

The spacetime dragging of inertial frames is one of the best examples of how the Mach principle and General Relativity (GR) are connected \cite{1992mtbh.book.....C}. A rotating BH can be revealed by its peculiar gravitational lensing connected with the gravitational Faraday effect: linearly polarized electromagnetic radiation propagating in a curved spacetime experiences a rotation of the polarization very similar to the Faraday rotation that occurs when the light is traversing a medium distorted by the presence of a magnetic field \cite{1973IJTP7467D}. Recently, it was shown that rotating black holes cause an additional effect on the phase of light the lensed that introduce an optical vorticity that depend on the rotation parameter of the BH \cite{tamburini2011twisting}.
The presence of the BH electric charge $Q$ has the effect of changing and, in certain cases, to enhance some of the effects expected in Kerr spacetimes \cite{1996AuJPh..49..623M}.
Proposals of detecting the electrostatic charge of black holes and, more precisely, of the central black hole in our galaxy have been widely discussed in the literature (see e.g. Ref. \cite{2005A&A...442..795Z} and the references therein).

In natural units, $e = G = c = 1$, Kerr-Newman space-times in the Boyer-Lindquist coordinates, $(t, r, \theta, \varphi)$, are described by the following line element,
\begin{eqnarray}
ds^2 & =&  - \left( {1 - \frac{{2Mr - Q^2 }}{{\rho ^2 }}} \right)dt^2  - \frac{{2a\left( {2Mr - Q^2 } \right)}}{{\rho ^2 }}\sin ^2 \theta d\varphi dt  \nonumber
\\ 
&+& \sin ^2 \theta \left( {r^2  + a^2  + \frac{{a^2 \left( {2Mr - Q^2 } \right)}}{{\rho ^2 }}\sin ^2 \theta } \right)d\varphi ^2  \nonumber
\\
&+& \frac{{r{}^2 + a^2 \cos ^2 \theta }}{\Delta }dr^2  + \rho ^2 d\theta ^2 
 \end{eqnarray}
where the quantities $\rho ^2  = r^2  + a^2 \,\cos ^2 \theta$ and $\Delta  = r^2  - 2\,Mr + a^2  + Q^2$
depend on mass $M$, electrostatic charge $Q$ and angular momentum per unit mass, $a$, also known as the black hole rotation parameter. For a Kerr-Newman spacetime, all the three parameters satisfy the rotation/charge/mass inequality, $a^2  + Q^2  \le M^2$. The presence of a de Sitter term describing a cosmological constant, $\Lambda$, will be discussed only in the non-rotating scenario.

Photons emitted or passing nearby such black holes are expected to acquire, because of the presence of turbulent plasma structures and, eventually, of gravitational lensing, an effective mass and orbital angular momentum that acts as a mass reducing  term \cite{refId}.

In fact, the acquisition of an effective mass by photons propagating in a plasma is described by the Anderson-Higgs mechanism \cite{PhysRev.130.439, PhysRev.125.397} or described in terms of an hidden gauge invariance of the Proca-Maxwell equations preserving Lorentz invariance  \cite{Mendonca:Book:2001}. Photons can acquire an additional virtual mass term in the presence of a cosmological charge asymmetry \cite{Dolgov200797}. The photon mass is supposed to be acquired through the interaction of the photons with a structured plasma \cite{refId} surrounding the compact object and the presence of massive photons is expected to modify the spacetime curvature around the BH \cite{procaRN,springerlink:10.1023/B:IJTP.0000048638.90043.97}.
For the sake of simplicity and to give better evidence to the r\^ole of the photon mass, we do not express directly in the line element the small gravitational perturbation caused by the presence of the plasma. 

In vacuum, photons are zero-rest mass particles and carry energy, momentum and angular momentum. The total angular momentum, $\mathbf{J}$, is given by the sum of the spin $\mathbf{S}$ and orbital angular momentum (OAM) $\mathbf{L}$ that can be transported to infinity \cite{Jackson:490457,Thide:Book:2009}. 
In the simplest case, when the plasma is homogeneous and characterized by its unperturbed plasma frequency $\omega_{p0}$, and when the perturbation induced by the photons in the plasma  is also neglected, photons will acquire a mass $m^{\ast} =\hbar\omega_{p0}/c^2$, where $\hbar$ and $c$ are the reduced Planck's constant and the speed of light, respectively. In natural units, this virtual photon mass coincides with the unperturbed plasma frequency.

\section{Kerr-Newman BH with Proca photon mass}

Let us consider a Kerr-Newman BH surrounded by an astrophysical turbulent plasma and, in the first approximation, neglect the modification of the spacetime curvature caused by the effect of plasma perturbations. The acquisition of a small mass term due to the interaction of photons in a plasma around a black hole will require a modification of Einstein's equations because the Proca stress-energy tensor is not traceless like that of Maxwell equations.

In this case, the description of electromagnetic phenomena given by Maxwell's equations are replaced by Proca equations
\begin{equation}
\nabla_\mu F^{\mu \nu} + \mu^2_\gamma = 4 \pi J^\nu
\end{equation}
where $\mu_\gamma$ is the general Proca photon mass, $\nabla_\mu$ is the covariant derivative,  $F^{\mu \nu}$ is the Maxwell tensor and $J^\nu$ the 4-current.

If the plasma is turbulent or has a characteristic spatial structure identified by a parameter $q$, it may induce also an orbital angular momentum $\ell$ to photons \cite{refId}. This term will directly appear in the Proca mass formulation,
\begin{equation}
\mu_\gamma =  \omega_{p0} \, A\, \sqrt{1- \frac{4 \pi \delta \dot{v}}{E+\hat{v} \cdot \nabla \phi} \frac{n_0+\tilde{n}f(\ell,q)}{{ \omega_{p0}^2 \, A}^2}}.
\label{mugamma}
\end{equation}
Here, $\hat{v}= \mathrm{\textbf{v}} /|v|$ is the direction vector of the electron velocity in the plasma, $\mathrm{\textbf{v}}$, which is, according to Ref. \cite{refId}, considered to be parallel to the electric field, $\mathrm{\textbf{E}}$. The vector potential of the EM field is $A$ and $\delta \dot{ v}=\hat v\cdot\partial_t v=|\partial_t v|$. 
The quantity $n_0$ is the unperturbed electron density and $m_e$ the electron mass; the term $\tilde n$ represents the so-called ``helicoidal'' density perturbation in the plasma that is described by $f(\ell,q)$ \cite{MajoranaTower}. During the propagation of a light beam only the total angular momentum,  $\mathbf{J}$, is preserved. When photons are traversing a structured inhomogeneous medium, plasma inhomogeneity applies 
torque to the photons and generate beams of photons carrying OAM \cite{2040-8986-13-6-064001}.

Massive photons will introduce in the original metric a Yukawa potential that depends on the electrostatic charge of the BH, $Q$, via the term $\Xi(Q)$ describing the charge asymmetry of the plasma distribution due to the black hole selective capture. For the trivial model of charged BH, $\Xi(Q)=Q^2/4 \pi$.

The modified Kerr-Newman metric with massive photons then becomes
\\
\begin{eqnarray}
&&ds^2=-\left(1-\frac{2Mr - \Xi(Q) e^{-\mu_\gamma  r}}{\rho^2}+\mu_\gamma  \int_r^\infty \Xi(Q)  \frac{e^{-\mu_\gamma  r}}{\rho^2}dr\right)dt^2 \nonumber
\\
&+&\left(\frac{r^2+a^2}{\rho^2} - \frac{2Mr - \Xi(Q) e^{-\mu_\gamma  r}}{\rho^2} + \mu_\gamma  \int_r^\infty  \Xi(Q) \frac{e^{-\mu_\gamma  r}}{\rho^2}dr \right)^{-1}dr^2 \nonumber
\\
&+& \rho^2 d\theta^2 + \left[ \rho^2 \sin^2 \theta + \left(\frac{2Mr -  \Xi(Q) e^{-\mu_\gamma  r}}{\rho^2} + \right. \right. \nonumber
\\
&+&  \left. \left.\mu_\gamma  \int_r^\infty  \Xi(Q) \frac{e^{-\mu_\gamma  r}}{\rho^2}dr \right) a^2 \sin^2 \theta \right] d\varphi^2 + 2a \sin^2 \theta \times \nonumber
\\
&&\times \left(\frac{2Mr - \Xi(Q) e^{-\mu_\gamma  r}}{\rho^2}- \mu_\gamma  \int_r^\infty \Xi(Q) \frac{e^{-\mu_\gamma  r}}{\rho^2}dr\right) d\varphi dt
\end{eqnarray}
where $\rho^2=r^2+a^2\cos^2 \theta$.

Let us consider the simplest case, when an helicoidal perturbation in the plasma electron distribution is present  In this case one has a direct correspondence between the spatial distribution and the orbital angular momentum acquired by photons in the Proca photon mass by deriving the OAM term $\ell$ from the approximation of the Proca mass of Ref. \cite{refId} to give a better evidence to the r\^ole of the orbital angular momentum in the spacetime curvature,
\begin{eqnarray}
&&\mu_\gamma^2 = \frac{E}{E+\hat{v}\cdot\nabla\phi}\omega^2_{p0}(1+\varepsilon) 
\\
&&-\frac{1}{E+\hat{v}\cdot\nabla\phi}\Big( 4\pi \delta \dot{ v}\big[n_0+\tilde{n}\cos(\ell \varphi^\ast+q\, r)\big]- 4 \pi \hat{v} \cdot \dalembert  \nabla \phi \Big) \nonumber
\label{eq:2}
\end{eqnarray}
here $\epsilon$ is the perturbation in the electron distribution and the term $q$ is the helix step of the ``helicoidal'' density perturbation in the plasma and $\varphi^\ast$ the azimuthal angle measured in the plane orthogonal to that of the radial coordinate in a local reference frame. Eq. \ref{eq:2} shows that the photon orbital angular momentum, $\ell$, acts as a \textit{negative squared mass term} for the Proca photon mass that clearly reduces the virtual photon mass.

The distorted geometry of a rotating BH induces OAM to the massive photons passing nearby independent of the frequency, with the result of modifying the spacetime curvature. At the first order, the metric will depend on the absolute value of the OAM acquired,
namely
\begin{equation}
\mu_\gamma \simeq P_\mu - \ell \;  \;  \frac{D_\mu \tilde{n} \sin(q \, r)}{2 \: P_\mu}
\label{eq6}
\end{equation}
where the constants characterizing the Proca mass are so defined
\begin{equation}
P_\mu = \sqrt{B_\mu+C_\mu  - D_\mu \cos(q \, r)}
\end{equation}
and
\\
\begin{eqnarray}
&B_\mu = \frac{E}{E+\hat{v}\bdot\grad\phi}\omega^2_{p0}(1+\varepsilon)
\\
&C_\mu = \frac{1}{E+\hat{v}\bdot\grad\phi}\Big( 4\pi \delta \dot{ v}n_0- 4 \pi \hat{v}\cdot\dalembert\nabla\phi\Big)
\\
&D_\mu =\frac{ 4\pi \varphi^\ast \delta \dot{ v}}{E+\hat{v}\bdot\grad\phi}.
\label{eq7}
\end{eqnarray}
If the plasma density tends to zero, the photon mass, $\mu \rightarrow 0$ and the Kerr-Newman spacetime  is recovered \cite{springerlink:10.1023/B:IJTP.0000048638.90043.97}.

The interplay between the photon Proca mass and the orbital angular momentum of light is known to be valid also at the quantum level and shows that the sign of the norm of the metric tensor, $||g||$, is preserved and remains positive-definite for any OAM value of the photon. Following the Majorana-Oppenheimer formulation of quantum electrodynamics \cite{2008PhRvA..78e2116T, giannetto85} in the Riemann-Silberstein formalism, the photon wavefunction $\mathrm{\textbf{G}}= \textbf{E} \pm  \mathrm{i} \textbf{B}$ \cite{Silberstein:AP:1907a, Silberstein:AP:1907b, Berry:JOA:2004a, Thide:Book:2009} obeys a Dirac-like equation with the well-known problems of  photon localizability \cite{mignani73,Birula:APP:1994, bialynicki1998exponential, Thide:Book:2009}. 
The spectrum of these  photon states exhibits a relationship between the total angular momentum, $J$, with the ``Majorana-mass'', $m^\ast$, related to the virtual mass acquired in the unperturbed plasma \cite{MajoranaTower},
\begin{equation}
\mu_\gamma =  \frac{m^\ast}{J(\ell,q)+ 1/2 }.
\label{eq8}
\end{equation}
Here, the term $J(\ell,q)$ is a general function of the orbital angular momentum of the photon and of the characteristic spatial scale of the perturbation. This relationship describes a spectrum cast in an infinite tower of photon states with positive-definite mass values that decrease when the total angular momentum of the particle increases: the higher  the value of the orbital angular momentum acquired by the photon, the lower the total effective Proca mass becomes, avoiding the photon to assume arbitrary  negative and negative unbounded squared mass values and preserving the signature of the metric tensor norm.

Differently from the mathematical structure of  the space-time manifold described by the Lorentz group, in which space is homogeneous and isotropic and time homogeneous, a plasma may exhibit peculiar spatial/temporal structures that breaks the space-time symmetry. The inhomogeneity of plasma structures and the interaction between photons and charges in the plasma are actually thought to alter the value of the Proca mass. This mass term, connected with time invariance, exhibits an additional more complicated space/time invariance that generates the conversion of a fraction of the Proca mass into photon orbital angular momentum. This phenomenon is independent of the acquisition of OAM generated by the gravitational field of a rotating BH.
In this latter case, the effect of the structures due to plasma turbulence is that of changing the value and even the signature of the photon OAM.
Moreover, the mathematical correspondence between the Majorana solution and the behavior of a photon in a plasma confirms the \textit{ansatz} that OAM states cannot induce a negative squared mass to photons, but a variation of it will relax the Yukawa potential of the spacetime curvature.

\section{Reissner-Nordstr\"om-de Sitter spacetime}
Let us extend this result to spacetimes that may be eligible to describe a cosmological scenario.
Spacetimes with a positive cosmological constant $\Lambda$ are supposed to be preferred for the description of the geometry of the Universe at large scale. Amongst the possible solutions, including that of a G\"odel universe, related to the Kerr-Newman case already discussed, we now focus our analysis to those models of spacetime that describe non-rotating universes. 
An example is the de Sitter metric or the Reissner-Nordstr\"om-de Sitter (RNdS) spacetime that describes a geometric support in the presence of electric charge. This class of solution can be eligible to provide a rough description of a Universe with accelerated expansion and a non-zero electric charge. Possible asymmetries in the total cosmological electric charge is supposed to be null or negligible, even if experimental data allow for possible small deviations without modifying Maxwell's electrodynamics.
The general RNdS line element is given by the quadratic form
\begin{eqnarray}
&&ds^2= - \left(1-\frac{2M}r - \frac \Lambda 3 r^2 + \nonumber
\frac{Q^2}{r^2}\right)dt^2 +
\\
&&+ \frac{dr^2}{\left(1- \frac{2M}r - \frac \Lambda 3 r^2+ \frac{Q^2}{r^2}\right)}+ r^2 \left(d\theta^2 + sin^2 \theta \d \varphi^2 \right)
\end{eqnarray}

If the plasma is located far away from charged bodies, $2M/r \ll 1$ and $\Lambda r^2 \ll 1$, considering the contribution of the Proca photon mass $\mu_\gamma$ and neglecting the second order quantities, the metrics becomes \cite{procaRN}
\begin{eqnarray}
&ds^2=-\Bigg(1-\frac{2M}{r} -\frac13\Lambda r^2 + \frac{Q^2}{r^2} e^{-2\mu_\gamma r} - \frac{2\mu_\gamma Q^2}r e^{-2 \mu_\gamma r}
\\
&- 4 \mu_\gamma^2 Q^2 \int^\infty_r \frac{e^{-2 \mu_\gamma r}}{r} dr \Bigg) dt^2 + \Bigg(1- \frac{2M}{r} - \frac 13 \Lambda r^2 + \frac{Q^2}{r^2} e^{-2 \mu_\gamma r} \nonumber
\\
&- \frac{2 \mu_\gamma Q^2}r e^{-2\mu_\gamma r} + \mu^2_\gamma Q^2 e^{-2 \mu_\gamma r} + 2r \mu^3_\gamma Q^2 e^{-2 \mu_\gamma r} \Bigg)^{-1} d r^2 \nonumber
\\
&+r^2 \Bigg(1+\mu^2_\gamma e^{-2 \mu_\gamma r} + 2 \mu^2_\gamma Q^2  \int^\infty_r \frac{e^{-2 \mu_\gamma r}}{r} dr \Bigg)  \times (d \theta ^2 + \sin^2 \theta d\varphi^2). \nonumber
\label{rnp}
\end{eqnarray}

By including the mass-to-OAM conversion described in Eq. \ref{eq6} and in Eq. \ref{eq8}, one can show that in the presence of a characteristic spatial scale and electron distribution given by the turbulence of plasma, the Proca photon mass will decrease the less the matter distribution in the Universe is homogeneous and isotropic. The line element will converge to that of the trivial Reissner-N\"ordstrom de Sitter spacetime.
We now expand the metric coefficient relative to the radial coordinate, $dr^2$, in the line element of Eq \ref{rnp} at the second order in $\ell$. In this way better evidence is given to the coupling between the OAM, acting as a negative mass term, with the cosmological constant $\Lambda$ and the electrostatic charge $Q$,

\begin{eqnarray}
&&\left(1+\frac{Q^2}{r^2}-\frac{2 M}{r} -\frac{\Lambda r^2}{3}\right)^{-1} 
\\
&&+\frac{36 Q^2 r^3 \left( P_\mu - \ell \;  \;  D_\mu \tilde{n} \sin(q \, r) /2 P_\mu
\right)}{\left(3 Q^2-6 M r+3 r^2-\Lambda r^4\right)^2} \nonumber
\\
&&+\frac{9 r^4 \left(27 Q^4+42 M Q^2 r-21 Q^2 r^2+7 \Lambda Q^2 r^4\right)}{\left(3 Q^2-6 M r+3 r^2-\Lambda r^4\right)^3  }\times \nonumber
\\
&&\times \frac{ [P_\mu - \ell \;  \;  D_\mu \tilde{n} \sin(q \, r) ]^2}{4 \, P_\mu^2} \sim O\left( \frac1{\Lambda r^2 }\right) +O\left( \frac {\ell \, Q^2}{\Lambda^2 r^4 } \right) \nonumber
\\
&&+O\left( \frac{-\ell^2  Q^2}{\Lambda^2 r^4 }\right)+ O\left( \frac {-\ell \, Q^2}{\Lambda^2 r^5 } \right). \nonumber
\label{pow}
\end{eqnarray}

What becomes immediately evident is that the coupling between $\ell$ and $\Lambda$ is very weak at large distances from the centre of mass and the higher power terms turn out to be independent, and thus invariant, of the BH mass. The most relevant OAM term in the power expansion of Eq. \ref{pow}, namely, $-\ell^2  Q^2 / \Lambda^2 r^4$, acts as an effective reducing mass term for the photon mass which is competing with the other term $\ell \, Q^2/\Lambda^2 r^4$ in case of negative values of $\ell$. For the specific case of $\ell=1$ the two terms tend to compensate each other, with the result that only terms with higher powers of $r$ act as  negative mass terms. Thus, OAM becomes a subtracting mass mechanism for massive photons, perceived as a missing mass in Einstein's equations that may induce an additional acceleration to that of the positive cosmological constant.
Of course, this effect cannot be held responsible for the acceleration of the expansion observed with supernova data \cite{mukhanov2005physical} or be concomitant with the hypothesized electrostatic repulsion of galaxies discussed in Ref. \cite{Dolgov200797}. 

\section{Conclusions}

Photons propagating in a structured plasma acquire mass and orbital angular momentum because of the hidden gauge invariance due to the Anderson-Higgs mechanism in Proca-Maxwell equations. 
In the presence of a plasma, the breaking of spatial homogeneity and a characteristic scale length $q$ introduced by the plasma structure, induce a mass/total angular momentum relationship with the properties of the spin/mass relationship expected for Majorana particles.
Because of this, the squared-mass term that describes photon OAM cause photons to acquire a smaller positive squared mass the higher the OAM value preserving the signature of the metric tensor.

The change of OAM states in the spacetime metric are evident only at very short distances from the black hole, dramatically modifying and flattening the Yukawa potential expected from the presence of massive photons for high values of OAM. The interplay between BH rotation, the acquisition of photon OAM from spacetime and from the plasma surrounding the BH will be revealed in the spiral spectrum of the photons emitted nearby the charge, mass, rotation relationship of a Kerr-Newman spacetime.
Moreover, in the presence of a negative cosmological constant, it becomes evident that the OAM term may concur with the cosmological constant to an accelerated expansion because of the decrease of the Proca photon mass.

\section*{References}

\begin{thebibliography}{10}
\expandafter\ifx\csname url\endcsname\relax
  \def\url#1{{\tt #1}}\fi
\expandafter\ifx\csname urlprefix\endcsname\relax\def\urlprefix{URL }\fi
\providecommand{\eprint}[2][]{\url{#2}}

\bibitem{1992mtbh.book.....C}
{Chandrasekhar} S 1992 {\em {The mathematical theory of black holes}\/} (New
  York: Oxford University Press)

\bibitem{Dolgov200797}
Dolgov A and Pelliccia D~N 2007 {\em Physics Letters B\/} {\bf 650} 97 -- 102

\bibitem{PhysRevD.66.024010}
Eiroa E~F, Romero G~E and Torres D~F 2002 {\em Phys. Rev. D\/} {\bf 66} 024010

\bibitem{1973IJTP7467D}
{Dehnen} H 1973 {\em International Journal of Theoretical Physics\/} {\bf 7}
  467--474

\bibitem{tamburini2011twisting}
Tamburini F, Thid{\'e} B, Molina-Terriza G and Anzolin G 2011 {\em Nature
  Physics\/} {\bf 7} 195--197 ISSN 1745-2473

\bibitem{1996AuJPh..49..623M}
{Metzenthen} W~E 1996 {\em Australian Journal of Physics\/} {\bf 49} 623--631

\bibitem{2005A&A...442..795Z}
{Zakharov} A~F, {de Paolis} F, {Ingrosso} G and {Nucita} A~A 2005 {\em Astron.
  Astroph.\/} {\bf 442} 795--799

\bibitem{refId}
{Tamburini, F}, {Sponselli, A}, {Thid\'e, B} and {Mendon\c{c}a, J T} 2010 {\em
  EPL\/} {\bf 90} 45001

\bibitem{PhysRev.130.439}
Anderson P~W 1963 {\em Phys. Rev.\/} {\bf 130} 439--442

\bibitem{PhysRev.125.397}
Schwinger J 1962 {\em Phys. Rev.\/} {\bf 125} 397--398

\bibitem{Mendonca:Book:2001}
Mendon\c{c}a J~T 2001 {\em Theory of Photon Acceleration\/} (Bristol, UK: IOP
  Publishing) {ISBN}~0-7503-0711-0

\bibitem{procaRN}
Shi C and Liu Z 2005 {\em International Journal of Theoretical Physics\/} {\bf
  44}(3) 303--308

\bibitem{springerlink:10.1023/B:IJTP.0000048638.90043.97}
Bei X, Shi C and Liu Z 2004 {\em International Journal of Theoretical
  Physics\/} {\bf 43}(6) 1555--1560

\bibitem{Jackson:490457}
Jackson J~D 1999 {\em Classical Electrodynamics\/} 3rd ed (New York, NY: Wiley
  \& Sons)

\bibitem{Thide:Book:2009}
Thid{\'e} B {\em {E}lectromagnetic {F}ield {T}heory\/} 2nd ed (New York, NY,
  \emph{in press}: Dover Publications)

\bibitem{MajoranaTower}
{Tamburini} F and {Thid\'e} B 2011  (\textit{Preprint}
  \eprint{arXiv:1105.0700})

\bibitem{2040-8986-13-6-064001}
Marrucci L, Karimi E, Slussarenko S, Piccirillo B, Santamato E, Nagali E and
  Sciarrino F 2011 {\em Journal of Optics\/} {\bf 13} 064001

\bibitem{2008PhRvA..78e2116T}
{Tamburini} F and {Vicino} D 2008 {\em Phys.\ Rev.\ A\/} {\bf 78} 052116--+
  (\textit{Preprint} \eprint{0807.3902})

\bibitem{giannetto85}
Giannetto E 1985 {\em Lettere Al Nuovo Cimento (1971--1985)\/} {\bf 44}
  140--144

\bibitem{Silberstein:AP:1907a}
Silberstein L 1907 {\em Ann.\ Phys. (Leipzig)\/} {\bf 327} 579--586

\bibitem{Silberstein:AP:1907b}
Silberstein L 1907 {\em Ann.\ Phys. (Leipzig)\/} {\bf 329} 783--784

\bibitem{Berry:JOA:2004a}
Berry M~V 2004 {\em J.~Opt.\ A: Pure Appl.\ Opt.\/} {\bf 6} S175--S177

\bibitem{mignani73}
R~Mignani E~R and Baldo M 1974 {\em Lett. Nuovo Cimento\/} {\bf 11} 568

\bibitem{Birula:APP:1994}
Bialynicki-Birula I 1994 {\em Acta\ Phys.\ Polon.\/} {\bf A 86} 97--116

\bibitem{bialynicki1998exponential}
Bialynicki-Birula I 1998 {\em Physical review letters\/} {\bf 80} 5247--5250
  ISSN 1079-7114

\bibitem{mukhanov2005physical}
Mukhanov V 2005 {\em Physical foundations of cosmology\/} (Cambridge Univ Pr)

\end{thebibliography}

\providecommand{\newblock}{}

\end{document}